\documentclass[notitlepage, 12pt, showkeys]{revtex4-1}
\usepackage[T1]{fontenc}
\usepackage{amssymb}
\usepackage{graphicx}
\usepackage{color}
\usepackage{caption}
\usepackage{hyperref}
\usepackage{tikz}

\usepackage{amsmath}
\usepackage{braket}
\usepackage{graphicx}
\usepackage{appendix}
\usepackage{subcaption}

\begin{document}
\title{Finite time path field theory and a new type of universal quantum spin chain quench behavior}

\author{Domagoj Kui\'{c}}

\email{Correspondence: dkuic@tvz.hr}

\affiliation{Zagreb University of Applied Sciences, Vrbik 8, 10000 Zagreb, Croatia}

\author{Alemka Knapp}

\email{aknapp@tvz.hr}

\affiliation{Zagreb University of Applied Sciences, Vrbik 8, 10000 Zagreb, Croatia}

\author{Diana \v{S}aponja-Milutinovi\'{c}}

\email{dsaponjam@tvz.hr}

\affiliation{Zagreb University of Applied Sciences, Vrbik 8, 10000 Zagreb, Croatia}

\date{July 24, 2025.}

\begin{abstract}
We discuss different quench protocols for Ising and XY spin chains in a transverse magnetic field. With a sudden local magnetic field quench as a starting point, we generalize our approach to a large class of local non-sudden quenches. Using finite time path field theory (FTPFT) perturbative methods, we show that the difference between the sudden quench and a class of quenches with non-sudden switching on the perturbation vanishes exponentially with time, apart from non-substantial modifications that are systematically accounted for. As the consequence of causality and analytic properties of functions describing the discussed class of quenches, this is true at any order of perturbation expansion and thus for the resummed perturbation series. The only requirements on functions describing the perturbation strength switched on at a finite time $t=0$ are as follows: (1) their Fourier transform $f(p)$ is a function that is analytic everywhere in the lower complex semiplane, except at the simple pole at $p=0$ and possibly others with $\Im (p) < 0$; and (2) $f(p)/p$ converges to zero at infinity in the lower complex semiplane. A prototypical function of this class is $\tanh(\eta t)$, to which the perturbation strength is proportional after the switching at time $t=0$. In the limit of large $\eta$, such a perturbation approaches the case of a sudden quench. It is shown that, because of this new type of universal behavior of Loschmidt echo (LE) that emerges in an exponentially short time scale, our previous results (Kui\'{c}, D. et al. \emph{Universe} \textbf{2024}, \emph{10}, 384) for the sudden local magnetic field quench of Ising and XY chains, obtained by the resummation of the perturbative expansion, extend in the long-time limit to all non-sudden quench protocols in this class, with non-substantial modifications systematically taken into account. We also show that analogous universal behavior exists in disorder quenches, and ultimately global ones. LE is directly connected to the work probability distribution, and the described universal behavior is therefore appropriate in potential concepts of quantum technology related to spin chains.
\end{abstract}

\keywords{finite time path field theory; quenches; Loschmidt echo; spin chains; perturbative methods}

\maketitle

\section{Introduction}
The question of dependence of Loschmidt echo (LE) and the intimately related statistics of work done in different systems, on the exact details of protocols leading to the transformation from the initial to a final state, is interesting from a wide variety of different aspects. In this regard, the appearance of universal dynamical behavior near quantum critical points and the issue of thermalization are probably among the most studied aspects (see for example \cite{Polkovnikov1, Mitra2, Das1} for reviews). Most of these studies apply either adiabatic (or nearly adiabatic), or on the opposite side, an instantaneous perturbation of the system known as a quench, achieved by a sudden change of one (or more) parameters of the Hamiltonian.

In relation to spin chains, quenches are utilized to elucidate a variety of different issues related to their nonequilibrium dynamics as follows: nonthermalization of integrable systems, describable using Generalized Gibbs Ensembles (GGE) \cite{Essler1}, long-lived prethermal states close to integrability~\cite{Mitra1, Marcuzzi1, Bertini1, Mitra2}, failure to thermalize in interacting systems with strong disorder and many body localization \cite{Nadkishore1, Yang1}, dynamical transitions not connected to equilibrium ones~\cite{Zunkovic1, Jafari1}, dynamical behavior near a critical point~\cite{Senegupta1, Guo1, Foini1, Paul1} and topological properties~\mbox{\cite{Ding1, Porta1}}, interplay between nonintegrability and integrability in the nonequilibrium response of the system~\cite{Lupo1}, nonequilibrium work probability and dynamical behavior near the critical point \cite{Silva1, Gambassi1, Smacchia1}, the effect of frustration in antiferromagnetic (AFM) spin chains with odd number of sites and periodic boundary conditions (topological frustration) \cite{groupLE, Catalano1}, and various other interesting and related aspects \cite{Fagotti1, Rossini1, Rossini2, Campos1, Canovi1, Campos3, Calabrese1, Igloi1, Igloi2, Riegler1, Schuricht1, Calabrese2, Calabrese3, Fagotti2, Heyl1}. With proposals~\cite{Rossini3, Goold1, Knap1, Knap2, Dora1, Dorner1, Mazzola1} for measurements of LE, such studies are interesting also from an experimental \mbox{point of view.}

In our recent paper \cite{Kuic1} we applied methods of finite time path field theory (FTPF)~\cite{Dadic1,Dadic2,Dadic3,Dadic4} to the calculation of the perturbative expansion of Loschmidt echo (LE) for the local magnetic field quenches of the ground states of Ising and XY chains in a transverse magnetic field. The diagonalization of these models is carried out (see \cite{Lieb1, Fabio-book} for details) through a mapping to the second quantized noninteracting fermions (via Jordan--Wigner transformation (JWT) followed by a Bogoliubov rotation). Ground states of these chains, when nondegenerate, and this is true for finite chains in most cases, can be represented as a vacuum of \mbox{Bogoliubov fermions. }

Since the local magnetic field quench (as opposed to a global one) turns the problem at hand into a nonintegrable one, we resorted in \cite{Kuic1} to FTPFT perturbative methods based on the Wigner transforms (WTs) of projected two-point retarded 
Green's functions developed by \cite{Dadic1}. This led to a straightforward calculation of the perturbative expansion of LE at any order of the perturbation and then, after introducing certain simple analyticity assumptions, full resummation of the perturbative expansion.

In the continuation of the work on this problem initially set out in \cite{Kuic1}, we extend it and establish a more general connection between a large class of local magnetic field quench protocols that include a sudden quench as a special case. The class is defined by the two following requirements on the function $\delta h(t)$ that controls the time-dependent perturbation strength, or to be more precise, on its Fourier transform $f(p)$: it is required that $f(p)$ is analytic everywhere in the lower complex semiplane, except at the simple pole at $p = 0$ and possibly other simple poles with $\Im (p) < 0$, and that $f(p)/p$ converges to zero at infinity in the lower complex semiplane. 

This class includes a wide range of quench protocols, where the perturbation strength $\delta h (t)$ is switched on non-suddenly at finite time $t=0$, including the prototypical function $\delta h(t) = \delta h \tanh (\eta t)$. This function has two important limits as follows: an ``adiabatic'' limit $\eta \rightarrow 0^+$, where the perturbation is switched on infinitely slowly, requiring an infinite amount of time, and a sudden quench limit $\eta \rightarrow \infty$, with the instantaneous switching on at $t=0$, described by $\delta h(t) = \delta h \theta (t)$, i.e., a Heaviside step function.

As a consequence of causality, only retarded two-point functions appear at any order of perturbative expansion in ``bubble'' diagrams. Diagrams involve convolution products of retarded fermion functions projected to a finite interval of time evolution with perturbation strength $\delta h(t)$ at the vertices. The required properties of the functions $\delta h(t)$ for all functions in this class reduce the difference between the sudden quench diagrams and non-sudden quench diagrams to terms that vanish exponentially in the long-time limit. The only difference that remains in the long-time limit appears at the vertices with the greatest and lowest time. It is systematically accounted for, and it is non-substantial compared to the common long-time behavior of all quenches in this class. This common universal behavior in the long-time limit includes the sudden quench behavior of LE described \mbox{in \cite{Kuic1}} as its most substantial part.

With a rather straightforward modification at the vertices of ``bubble'' diagrams, and under the same requirements on the function $\delta h(t)$ that controls the switching on the perturbation, we show that completely analogous form of universal behavior exists for a disorder quench, which is described by random magnetic field perturbation variables all distributed in the interval $[-\delta h(t), \delta h(t)]$ along the entire chain. In fact, this particular modification also includes a special case, when the perturbation has the same value along the chain, which then implies an analogous universal behavior in global quenches also. 

Even though the results presented in \cite{Kuic1} and in this paper are reached using perturbative techniques, this universal behavior of LE emerging for large times seems to point out that on such large time scales spin chains are largely insensitive to the details of quench protocols. As is well known, complex conjugate of the LE amplitude is equal to the characteristic function of the work probability distribution \cite{Silva1}. Therefore, statistics of the work done on these systems (moments and cumulants of the work distribution) and of energy stored in a quench process is largely independent of the degree of experimental control and the details of the protocol. While there were previous suggestions that this may be the case near the critical point of the Ising chain \cite{Smacchia1}, our work presented here sets out some very general conditions on quench protocols that can possibly extend this arbitrary far from criticality.

The layout of the paper is as follows. In Section \ref{The model}, we revisit only the necessary details from \cite{Kuic1} about the AFM version of XY chain in a transverse magnetic field, a sudden local magnetic field quench, here applied at an arbitrary site of the chain, and about LE. These involve non-essential modifications in quench details w.r.t. to \cite{Kuic1} due to the arbitrariness of the quench site. In Sections \ref{perturbative_calculations} and \ref{projected_functions_WTs_convolutions} we give a basic explanation of the perturbative expansion of LE in terms of ``bubble'' diagrams, vertices, and basic concepts from FTPFT, projected retarded two-point fermion functions, their Wigner transforms (WTs), and their convolution products. We only present our final result from \cite{Kuic1} for the perturbation expansion up to an arbitrary order, and for the resummed perturbation series. This sudden quench result represents, up to non-substantial modifications, the common universal long-time behavior for a class of non-sudden quench protocols that we introduce in Section \ref{non-sudden_quenches}. In it, we analyze in great detail the conditions under which this universal behavior emerges in the long-time limit. In Section \ref{disorder_quenches}, we describe how, with minimal modifications in the resummed perturbative expansion of LE, an analogous universal behavior emerges in the long-time limit for disorder quenches and for global quenches. Conclusions of this work are set forth in Section \ref{conclusions}.

\section{Sudden Local Quench of the XY Chain and Loschmidt Echo} \label{The model}

The XY spin chain in a transverse magnetic field $h$ is described by the Hamiltonian 
\begin{equation}
H_0 =J \sum_{j=1}^N (\frac{1+\gamma}{2}\sigma_j^x \sigma_{j+1}^x +\frac{1-\gamma}{2}\sigma_j^y \sigma_{j+1}^y+ h\sigma_j^z) . \label{XY_Hamiltonian}
\end{equation}

Pauli spin operators $\sigma_j^\alpha$, with $\alpha =x, y, z$, describe spins at different sites of the chain, each denoted by $j$. Parameter $J \lessgtr 0$ corresponds to FM or AFM versions of the model. It is set here to $J = + 1$ without losing generality. Values of $xy$ plane anisotropy parameter $\gamma = \pm1$ correspond to the quantum Ising chain. PBCs are given by $\sigma_{j+N}^\alpha \equiv \sigma_{j}^\alpha$, where $N$ is the number of sites. This imposes translational symmetry on the chain which is equivalent to a ring geometry. The integrability of the Hamiltonian (\ref{XY_Hamiltonian}) and the method of diagonalization is well known; it is described, for example, in \cite{Lieb1, Fabio-book}, and also in \cite{Kuic1}. It consists firstly of mapping by JWT to a spinless noninteracting fermionic model on a 1D lattice. Then, this second quantized fermionic Hamiltonian is diagonalized by a discrete Fourier transform followed by a Bogoliubov transformation in momentum space (see \cite{Kuic1} further on for details and notation).

We consider as a starting point a sudden perturbation of the Hamiltonian $H_0 \rightarrow H_1 = H_0 + V$. It is carried out by an instantaneous change $\delta h$ of the magnetic field strength $h$ at spin site $j$, so that 
\begin{equation}
V = \delta h \sigma_j^z . \label{V_spins}
\end{equation}

Because of the translational symmetry of $H_0$, the particular choice of the site $j$ is irrelevant. Local perturbation given by (\ref{V_spins}) breaks the translational symmetry of the model (\ref{XY_Hamiltonian}) but not $\mathbb{Z}_2$ parity symmetry, with the parity operator $\Pi^z = \bigotimes_{j=1}^N \sigma_j^z$. This is a $\pi$ angle $z$-axis rotation symmetry operator, up to an irrelevant multiplicative factor $e^{-iN\frac{\pi}{2}}$, which depends on the total number of sites.

The Hamiltonian (\ref{XY_Hamiltonian}) has two sectors, each corresponding to an eigenspace of the parity operator $\Pi^z$ with an eigenvalue $+1$ or $-1$. In each parity sector, the Hamiltonian is diagonal in terms of Bogoliubov fermions with creation and annihilation operators $\eta _{q}^\dagger$ and $\eta _{q}$,
\begin{equation}
H^{\pm } = \sum_{q \in \Gamma ^{\pm }} \epsilon (q) \left (\eta _{q}^\dagger \eta _{q} - \frac{1}{2}\right) . \label{H_P-S_Bogoliubov}
\end{equation}

Sum in (\ref{H_P-S_Bogoliubov}) is over the set of fermion momenta $\Gamma^+ = \{q = 2\pi(k + \frac{1}{2})/N : k =0, 1, \dots, N - 1 \} $ or $\Gamma^- = \{q = 2\pi k /N : k =0, 1, \dots, N - 1 \}$, depending on the parity ($\pm$). Excitation energies of Bogoliubov fermions $\epsilon (q) = 2\Lambda_q$ are given by

\begin{equation}
\epsilon (q) = 2\Lambda _q = \left \{ \begin{array}{l@{\,,\quad }l} 2 \left [(h - \cos q)^2 +\gamma^2 \sin^2 q \right ]^{1/2} & \forall q \in \Gamma ^{\pm } \backslash \{0, \pi \} \\ 2(h - \cos q) & N \ \mathrm{even}, \{q = 0, \pi\} \in \Gamma ^{- } \\ - 2(h - \cos q) & N \ \mathrm{odd}, h < 0, \{q = 0\} \in \Gamma ^{- }, \{q=\pi\} \in \Gamma ^{+ } 
\\ 2(h - \cos q) & N \ \mathrm{odd}, h > 0, \{q = 0\} \in \Gamma ^{- }, \{q=\pi\} \in \Gamma ^{+ } \end{array} \right . . \label{excitation_energies}
\end{equation}

Expressed in terms of Bogoliubov operators, the perturbation (\ref{V_spins}) is given by
\begin{equation}
V = \frac{\delta h}{N} \sum_{q, q^\prime \in \Gamma ^\pm }e^{i(q-q^{\prime})j} \left( \begin{array}{cc} \eta _{q}^\dagger & \eta _{-q} \end{array} \right ) a_{q, q^\prime} \left( \begin{array}{c} \eta _{q^\prime} \\ \eta _{-q^\prime}^\dagger \end{array} \right ) . \label{V_Bogoliubov}
\end{equation}

This expression is written specifically for the case $N$ is even and $h \ne 0$, and also for the case $N$ is odd and $h > 0$. When $N$ is odd and $h < 0$, all expressions starting from this point are obtained by replacing $\delta h \rightarrow - \delta h$. Matrix $a_{q, q^\prime }$ depends on the Bogoliubov angles,
\begin{equation}
a_{q, q^\prime} = \left( \begin{array}{c c} \cos (\theta _q + \theta _{q^\prime}) & -i\sin (\theta _q + \theta _{q^\prime})\\ i\sin (\theta _q + \theta _{q^\prime})& - \cos (\theta _q + \theta _{q^\prime}) \end{array} \right ) . \label{V_matrix}
\end{equation}

Sums in (\ref{V_Bogoliubov}) are over the set of fermion momenta $\Gamma^+$ or $\Gamma^-$, depending on the parity of the state on which the perturbation acts. Angle $\theta _q$ of the Bogoliubov transformation, which completes the diagonalization of the Hamiltonian (\ref{XY_Hamiltonian}), depends on the number of sites $N$ and on parameters $h$ and $\gamma$, and it is implicitly expressed as,
\begin{equation}
e^{i2\theta _q} = \left \{ \begin{array}{l@{\,,\qquad }l} \frac{ h - \cos q - i\gamma \sin q}{\Lambda _q} & N \ \mathrm{even} \\ \frac{ - h + \cos q + i\gamma \sin q}{\Lambda _q} & N \ \mathrm{odd}, h < 0 
\\ \frac{h - \cos q - i\gamma \sin q}{\Lambda _q} & N \ \mathrm{odd}, h > 0 \end{array} \right . . \label{Bogoliubov_angle}
\end{equation}

For a system in a state $\ket{\psi }$ and a sudden quench of the Hamiltonian $H_0 \rightarrow H_1$ at $t=0$, LE is defined as $\mathcal{L}(t)= |\mathcal{G}(t)|^2$, with its complex amplitude 
\begin{equation}
\mathcal{G}(t) = \bra{{\psi }}e^{iH_0t}e^{-iH_1t}\ket{\psi } . \label{complex_amplitude_LE}
\end{equation}

The definition of $\mathcal{L}(t)$ can be given different interpretations (for example, see \cite{Kuic1}, and for more details \cite{Goussev1}). We mention here one more interpretation, based on the relation to the work probability distribution $P(W)$. Its characteristic function $G(t) = \int dW e^{iWt}P(W) $ can be identified with the complex conjugate of LE amplitude, $G(t) = \mathcal {G}^{\ast} (t)$. The details are given in \cite{Silva1}, we reproduce it here only for better understanding of this point. 

In a quench, the system is out of equilibrium and the process is not quasistatic. Work is therefore characterized by a probability distribution $P(W)$. Determining $P(W)$  requires subsequent measurements of energy, as $W = E_1 - E_0$, where $E_0$ is the energy measured before the quench, and $E_1$ after the quench. Measured values $E_{0,m}$ and $E_{1,n}$ are the energy eigenvalues of $H_0$ and $H_1$, respectively. Thus, the work probability distribution, as given in~\cite{Silva1}, is equal to
 \begin{equation}  
P(W) = \sum _{n,m} \delta [W - (E_{1, n } - E_{0, m})]  \left | \langle \Phi _{1, n} | \Psi_{0, m} \rangle \right |^2 P_m  . \label{work_probability distribution}
\end{equation}

Here, $| \Psi_{0, m} \rangle$ and $| \Phi_{1, n} \rangle$ are the energy eigenstates of $H_0$ and $H_1$, respectively. $P_m$ is the initial statistical ensemble distribution. As $\delta () $ is a Dirac delta function, the characteristic function of $P(W)$ is equal to
\begin{equation}
G(t) = \int dW e^{iWt}P(W) =  \sum _{n,m} e^ {i(E_{1, n } - E_{0, m})t}  \left | \langle \Phi _{1, n} | \Psi_{0, m} \rangle \right |^2 P_m .  \label{work_probability distribution_char_f}
\end{equation}

It is clear that (\ref{work_probability distribution_char_f})  is a conjugate of LE complex amplitude, i.e.,  $G(t) = \mathcal {G}^{\ast} (t)$. In a case when the system is initially in an eigenstate $| \Psi_{0, m} \rangle $ of  $H_0$, or a in a superposition of $H_0$ eigenstates, LE complex amplitude  $\mathcal{G}(t)$ reduces to  the Formula (\ref{complex_amplitude_LE}).

\section{Perturbative Expansion} \label{perturbative_calculations}

The perturbative expansion of $\mathcal{G}(t)$ is based on applying the time evolution operator in the interaction picture $U_I(t,0)$. In this way, $\mathcal{G}(t)$ can be written for a generic time-dependent perturbation, including both sudden and non-sudden switching on the perturbation at $t=0$, 
\begin{eqnarray}
\mathcal{G}(t) & =& \bra{g_0}U_I(t,0)\ket{g_0} = \bra{g_0}Te^{-i\int_0^tV_I(t^\prime)dt^\prime}\ket{g_0} \nonumber \\
& = & \sum_{n=0}^\infty\frac{(-i)^n}{n!}\int_0^t dt_1 \dots \int_0^t dt_n \bra{g_0}T[V_I(t_1) \dots V_I(t_n)]\ket{g_0} . \label{G_perturbative}
\end{eqnarray}
$T$ is the time-ordered product of operators $V_I (t) = e^{iH_0t} V e^{-iH_0t}$. We first restrict the calculation of $\mathcal{G}(t)$ to the ground state $\ket{g_0}$ of the unperturbed Hamiltonian. 

We assume that the ground state is nondegenerate; this is true for finite $N$ if $h^2 >|1-\gamma^2|$ (see \cite{Kuic1,Fabio-book, Damski1} for further reference). For a finite size Ising chain ($\gamma = \pm 1$), the condition of ground state nondegeneracy reduces to $h \ne 0$. 

As described in \cite{Kuic1}, in all such cases, the ground state can be described as a vacuum of Bogoliubov fermions. When $N$ is even and $h \ne 0$, and also when $N$ is odd and $h < 0$, the ground state has positive parity $\Pi^z \ket{g_0^+} = \ket{g_0^+}$. When $N$ is odd and $h > 0$, the ground state is of negative parity $\Pi^z \ket{g_0^-} = -\ket{g_0^-}$. From (\ref{H_P-S_Bogoliubov}) we see that the energy of these ground states $H_0\ket{g_0^ \pm }=E_0^{GS \pm }\ket{g_0 ^\pm }$ is given by $E_0^{GS \pm } = -\sum_{q \in \Gamma^\pm }\epsilon (q)/2$.

In accordance with the linked-cluster theorem \cite{Coleman, Lancaster}, perturbative expansion (\ref{G_perturbative}) can be written as
\begin{eqnarray}
\mathcal{G}(t) = e^{\log[\mathcal{G}(t)]} = e^{ \sum_{n=1}^\infty\frac{(-i)^n}{n!}\int_0^t dt_1 \dots \int_0^t dt_n \bra{g_0}T[V_I(t_1) \dots V_I(t_n)]\ket{g_0}_C} , \label{G_cumulant_exp}
\end{eqnarray}
where $C$ denotes that the exponential contains a sum of only connected vacuum averages (or diagrams), obtained by applying Wick's theorem to the time-ordered products in (\ref{G_cumulant_exp}). Thus, the $n$-th order perturbation term in the exponent of (\ref{G_cumulant_exp}) contains only connected vacuum ``bubble'' diagrams with $n$ inserted vertices, with two strings of vacuum contractions joined at the greatest and least time vertices. This is all explained in much more detail in \cite{Kuic1}. 

Here we only briefly explain that, irrespective of the appearance of the additional factor $e^{i(q-q^{\prime})j}$ in (\ref{V_Bogoliubov}) and (\ref{V_matrix}) and in the vertex factors, as compared to the analogous expressions in \cite{Kuic1}, nothing else is substantially changed.

The greatest time or least time vertex factors are $ \pm i (\delta h /N)e^{i(q-q^{\prime})j} \sin (\theta _q + \theta _{q^\prime})$, respectively, and the intermediate time vertex factors are $\pm (\delta h /N)e^{i(q-q^{\prime})j} \cos (\theta _q + \theta _{q^\prime})$, as can be directly deduced from the form of perturbation matrix (\ref{V_Bogoliubov}) and (\ref{V_matrix}). Time ordering and contracting is responsible for all intermediate vertices obtaining a $+$ sign in front of vertex factor $(\delta h /N )e^{i(q-q^{\prime})j} \cos (\theta _q + \theta _{q^\prime})$ in a diagram. 

Vacuum contractions of Bogoliubov annihilation $\eta _{q}(t) = e^{iH_0 t} \eta _{q}e^{-iH_0 t} $ and creation operators $\eta _{q}^\dagger (t) = e^{iH_0 t} \eta _{q}^\dagger e^{-iH_0 t} $ appearing between vertex factors in these diagrams are retarded two-point Green's functions,
\begin{equation}
G_{R, q} (t_1-t_2)= \bra{g_0^{\pm}}\eta _{q}(t_1) \eta _{q}^\dagger(t_2) \ket{g_0^{\pm}}\theta(t_1 - t_2) = e^{-i 2\Lambda _{q} (t_1-t_2)}\theta(t_1 - t_2) . \label{Green_2point_retarded}
\end{equation}

Therefore, the WT of (\ref{Green_2point_retarded}), i.e., its Fourier transform with respect to $s_0 = t_1 - t_2$, is a function that has a pole in the lower complex semiplane,
\begin{equation}
G_{R,q}(p_0) = \frac{i}{p_0 - 2\Lambda _{q} + i\varepsilon} . \label{WT_Green_2point_retarded}
\end{equation}

For similar reasons, all ``bubble'' diagrams, ``direct'' and ``twisted'', appearing at the same order of perturbation have the same factor in front. A ``twist'', taken with respect to a ``direct'' diagram, means only an inversion of two Bogoliubov fermionic operators in the least time vertex when contracting them with the rest of the two strings. They are not topologically different but only represent different order of contractions in the least \mbox{time vertex. }

Equal time contractions are not included in (\ref{G_cumulant_exp}) because they generate disconnected diagrams.  There is a total of $2^{n-1} n! \sum_{k=0} ^{n-2} {n -2\choose n -2 - k}$ ``bubble'' diagrams appearing at the $n$-th order of perturbative expansion.  Factor $n!$ is due to the number of possible time orderings in the ``bubble'' diagrams of the same form at the $n$-th order of  (\ref{G_cumulant_exp}). Due to two possible types of intermediate time vertices, and a possibility of a ``twist'' in the lowest time vertex, this gives a total of $2^{n-1}n!$ ``bubble'' diagrams of the same form. 

Translational and other symmetries of the Hamiltonian (\ref{XY_Hamiltonian}) have the consequence that $\Lambda _q = \Lambda_{-q} = \Lambda_{N-q}$ and $\theta _q = - \theta_{-q} = - \theta_{N-q}$, as is evident from (\ref{excitation_energies}) and (\ref{Bogoliubov_angle}). This is used in showing that at the $n$-th order of  (\ref{G_cumulant_exp})  there is  $2^{n-1} n!$ ``bubble'' diagrams that have exactly the same form, which is given by a term in the sum over $j$ (i.e., the sum over all forms) in,
\begin{eqnarray}
&& \bra{g_0^{\pm}}T[V_I(t_1) \dots V_I(t_n)]\ket{g_0^{\pm}}_C \cr \nonumber\\ 
&& =  - 2^{n-1}(\delta h )^n \sum_{\mathrm{all \ time \ ord.}}\sum_{j=2}^{\mathrm{n}} \sum_{q_1, q_2 , \dots, q_n \in \Gamma ^{\pm }} \left \{\sin (\theta_{q_1} + \theta_{q_2})G_{R, q_2} (s_{0, 2})  \right . \cr \nonumber\\ 
&& \left . \times \cos (\theta_{q_2} + \theta_{q_3})G_{R, q_3} (s_{0, 3}) \cos (\theta_{q_3} + \theta_{q_4})  \cdots \cos (\theta_{q_{j-1}} + \theta_{q_{j}})G_{R, q_{j}} (s_{0, j})  \right . \cr \nonumber\\ 
&& \left .  \times \sin (\theta_{q_{j}} + \theta_{q_{j+1}})  G_{R, q_{j+1}}( s_{0, j+1})  \right . \cr \nonumber\\  
&& \left . \times  G_{R, q_{1}} (s_{0, 1})\cos (\theta_{q_{1}} + \theta_{q_{n}}) G_{R, q_{n}} (s_{0, n})\cos (\theta_{q_{n}} + \theta_{q_{n-1}}) \cdots \cos (\theta_{q_{j+2}} + \theta_{q_{j+1}})  \right \}  ,  
\label{n_ord_time_ordered product}
\end{eqnarray}
where $s_0$ are relative time variables. The first sum in (\ref{n_ord_time_ordered product}) is over all time orderings such that all $s_0 > 0$, because all $G_{R, q} (s_{0})$ are retarded two-point functions. Here, $\sin (\theta_{q_1} + \theta_{q_2})$ and $ \sin (\theta_{q_{j}} + \theta_{q_{j+1}})$ correspond to vertices with the greatest and lowest time, respectively.

At each order of perturbation expansion, time integrations in (\ref{G_cumulant_exp}) result in convolution products of retarded two-point functions (\ref{Green_2point_retarded}), with vertices coming from (\ref{V_Bogoliubov}) and (\ref{V_matrix}) inserted. Time integrals in (\ref{G_cumulant_exp}) thus make ``bubble'' diagrams, consisting of two strings of convolution products, joining at the vertex with the greatest time and at the vertex with the lowest time.

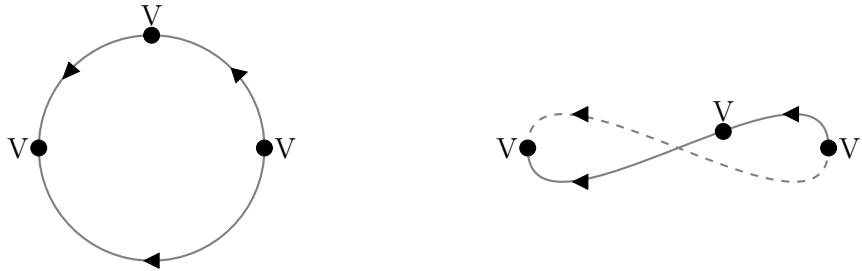
\begin{figure}[!h]
\begin{tikzpicture}
\draw[gray, thick] (0,0) circle (1.5);
\filldraw[black] (-1.5,0) circle (3pt) node[anchor=east] {V};
\filldraw[black] (1.5,0) circle (3pt) node[anchor=west] {V};
\filldraw[black] (0,1.5) circle (3pt) node[anchor=south] {V};

\draw [black, fill=black] (0.1,-1.6)--(-0.1,-1.5)--(0.1,-1.4)--(0.1,-1.6);
\draw [black, fill=black] (1.06,1.06)--(1.12,0.87)--(1.27,1.01)--(1.06,1.06);
\draw [black, fill=black] (-1.18, 0.92)--(-0.96,1.02)--(-1.1,1.14)--(-1.18,0.92);

\draw[gray, thick, dashed] (5,0) to [out=90,in=-90] (9,0); 
\draw[gray, thick] (5,0) to [out=-90,in=90] (9,0);
\filldraw[black](5,0) circle (3pt) node[anchor=east] {V};
\filldraw[black](9,0) circle (3pt) node[anchor=west] {V};

\filldraw[black] (7.6,0.22) circle (3pt) node[anchor=south] {V};

\draw [black, fill=black] (5.8,0.55)--(5.6,0.45)--(5.8,0.35)--(5.8,0.55);
\draw [black, fill=black] (5.8,-0.55)--(5.6,-0.45)--(5.8,-0.35)--(5.8,-0.55);
\draw [black, fill=black] (8.6,0.55)--(8.4,0.45)--(8.6,0.35)--(8.6,0.55);
\end{tikzpicture}
\caption{``Direct'' and ``twisted'' ``bubble'' diagrams (left and right part of the figure, respectively) appearing at the $3$rd order of (\ref{G_cumulant_exp}). Dashed line indicates that there is no intersection with a solid line. Arrows indicate the positive direction of time and of fermion propagation between vertices, which are denoted by points. V is an element of perturbation matrix (\ref{V_Bogoliubov}) and (\ref{V_matrix}) and a vertex in the two strings of vacuum contractions in the time-ordered product of fermionic operators comprising a ``bubble'' diagram.}\label{fig1}
\end{figure}

Types of ``bubble'' appearing at the third order term in perturbative expansion (\ref{G_cumulant_exp}) are depicted on Figure \ref{fig1}.

\vspace{-6pt}

\section{Projected Functions, WTs and Convolution Products} \label{projected_functions_WTs_convolutions}

In a typical field theory setup interactions are switched on adiabatically from $t= - \infty$, and time evolution follows accordingly. The quench-like protocol of Section (\ref{The model}) naturally fits within the formalism of FTPFT developed in \cite{Dadic1}; perturbation is switched on at $t=0$ and time evolution is then followed up until a finite time $t$. As a consequence, all time integrals in the convolution products of retarded two-point functions appearing in (\ref{G_cumulant_exp}) are over a finite interval (e.g., at $n$-th order over the region $0 \leq t_1, t_2, \dots, t_n \leq t$).

This is equivalent to all retarded two-point functions (\ref{Green_2point_retarded}) being projected to a finite time interval of evolution $0 \leq t_1, t_2 \leq t$,
\begin{equation}
G_{t, R, q} (t_1-t_2) = \theta(t) \theta(t - t_1) \theta(t_1) \theta(t - t_2)\theta(t_2) e^{-i 2\Lambda _{q} (t_1-t_2)}\theta(t_1 - t_2) . \label{Green_2point_retarded_projected}
\end{equation}

The product of $\theta$-functions $\theta(t) \theta(t - t_1) \theta(t_1) \theta(t - t_2)\theta(t_2)$ in front of $G_{R, q} (t_1-t_2)$ in (\ref{Green_2point_retarded_projected}) projects it to the interval $0 \leq t_1, t_2 \leq t$. The WTs of the projected functions have an important property, written here for $G_{t, R, q} (t_1-t_2)$, as follows: 
\begin{equation}
G_{t, R, q} (p_o) = \int_{-\infty}^{\infty} dp_0^\prime \frac{\theta(t)}{\pi}\frac{\sin(2t(p_0 - p_0^\prime))}{p_0 - p_0^\prime}G_{R,q}(p_o^\prime) , \label{WT_Green_2point_retarded_projected} 
\end{equation}
where $G_R(p_o)$ is the WT of (\ref{Green_2point_retarded}) given by (\ref{WT_Green_2point_retarded}). The factor in front of $G_R(p_o)$ in (\ref{WT_Green_2point_retarded_projected}) is the WT of the projector. Reference \cite{Dadic1} introduces projected functions, their WTs, and convolution products within the FTPFT framework. In our recent paper \cite{Kuic1} we have emphasized some properties of WTs of projected functions taken from \cite{Dadic1} that are important for this \mbox{work too. }

Here we use another important rule for the convolution products of projected function, also taken from \cite{Dadic1}. The convolution product of $n$ projected functions $\{A_{t, q} (t_1 - t_2) = \theta(t) \theta(t - t_1) \theta(t_1) \theta(t - t_2)\theta(t_2)A_{q} (t_1 - t_2) : q \in \Gamma ^{\pm}\}$, 
\begin{equation}
C_t (t_1 - t_{n+1}) = \int_{-\infty}^{\infty} dt_2 \dots \int_{-\infty}^{\infty} dt_{n}A_{t, q_1} (t_1-t_2) \cdots A_{t, q_n} (t_{n}-t_{n+1}) , \label{convolution_product_gen}
\end{equation}
has a WT given by the rule
\begin{eqnarray}
&& C_t (p_0) = \int_{-\infty}^{\infty} \left (\prod_{j=1}^{n} dp_{0,j} \right ) \frac{\theta(t)}{\pi}\frac{\sin\left [2t \left (p_0 - \frac{p_{0,1} + p_{0,n}}{2} \right )\right ]}{p_0 - \frac{p_{0,1} + p_{0,n}}{2} } \cr\nonumber \\
&& \times \prod_{j=1}^{n-1}\left (A_{q_j}(p_{0,j}) \frac{1}{2\pi}\frac{i}{p_{0,j} - p_{0, j +1} + i \varepsilon} \right ) e^{-it(p_{0,1} - p_{0,n} + i(n-1)\varepsilon)} A_{q_n}(p_{0,n}) . \label{WT_convolution_product_gen}
\end{eqnarray}
$A_{q_j}(p_{0,j})$ in (\ref{WT_convolution_product_gen}) is a WT of the function $A_{q_j} (t_1 - t_2) $. 

Now we consider the case that all functions in the convolution product (\ref{convolution_product_gen}) are projected retarded functions (\ref{Green_2point_retarded_projected}). Then, from (\ref{WT_convolution_product_gen}), by taking the $dp_{0,j}$ integrals in the order from the rightmost to the leftmost and closing the contours in the upper semiplane, \mbox{one gets}
\begin{eqnarray}
C_t (p_0) = \int_{-\infty}^{\infty} dp_0^\prime \frac{\theta(t)}{\pi}\frac{\sin(2t(p_0 - p_0^\prime))}{p_0 - p_0^\prime}G_{R,q_1}(p_o^\prime) \cdots G_{R,q_n}(p_o^\prime) . \label{WT_convolution_product}
\end{eqnarray}

Here, $G_{R,q_1}(p_o^\prime) \cdots G_{R,q_n}(p_o^\prime)$ is a WT of a convolution product of $n$ retarded functions~(\ref{Green_2point_retarded}). We see that convolution products of projected retarded functions~(\ref{Green_2point_retarded_projected}) obey the same WT rule as the functions themselves (see reference \cite{Dadic1} for more details about functions satisfying this rule).

Using the rule (\ref{WT_convolution_product}) we can calculate the two strings of convolution products of projected retarded two point functions $G_{t, R, q} (t_1-t_2)$ in between the greatest and lowest time vertices $\pm i (\delta h /N)e^{i(q-q^{\prime})j}\sin (\theta _q + \theta _{q^\prime})$ at which they join. The intermediate time vertices $ (\delta h /N)e^{i(q-q^{\prime})j}cos (\theta _q + \theta _{q^\prime})$ are inserted between the retarded two-point functions. At the $n$-th order of the exponent in (\ref{G_cumulant_exp}), using (\ref{n_ord_time_ordered product}), we obtain
\begin{eqnarray}
&& \frac{(-i)^n}{n!} \int_0^t dt_1 \dots \int_0^t dt_n \bra{g_0^{\pm}}T[V_I(t_1) \dots V_I(t_n)]\ket{g_0^{\pm}}_C \cr \nonumber \\
&& = \frac{(-i)^n}{n!} \int_{0}^t dt_1 \int_{0}^t dt_n \theta(t_1 - t_n) \left (- \frac{2^{n-1}n!(\delta h )^n i^n}{N^n}\right) \sum_{j=2}^{\mathrm{n}} \sum_{q_1, q_2 , \dots, q_n \in \Gamma ^{\pm }} \cr \nonumber\\ 
&& \left \{\sin (\theta_{q_1} + \theta_{q_2}) \left [ \frac{1}{2\pi}\int_{-\infty}^{\infty} dp_0 e^{-ip_o(t_1 - t_n)} \int_{-\infty}^{\infty} dp_0^\prime \frac{\theta(t)}{\pi}\frac{\sin(2t(p_0 - p_0^\prime))}{p_0 - p_0^\prime}\right . \right . \cr \nonumber\\
&& \left. \left . \frac{\cos (\theta_{q_2} + \theta_{q_3})\cos (\theta_{q_3} + \theta_{q_4}) \cdots \cos (\theta_{q_{j-1}} + \theta_{q_{j}})}{(p_0^\prime - 2\Lambda _{q_2} + i\varepsilon) (p_0^\prime - 2\Lambda _{q_3} + i\varepsilon) \cdots (p_0^\prime - 2\Lambda _{q_{j-1}} + i\varepsilon)(p_0^\prime -2 \Lambda _{q_j} + i\varepsilon)} \right ] \right . \cr \nonumber\\
&& \left . \sin (\theta_{q_{j}} + \theta_{q_{j+1}}) \left [ \frac{1}{2\pi}\int_{-\infty}^{\infty} dk_0 e^{-ik_o(t_1 - t_n)} \int_{-\infty}^{\infty} dk_0^\prime \frac{\theta(t)}{\pi}\frac{\sin(2t(k_0 - k_0^\prime))}{k_0 - k_0^\prime} \right . \right . \cr \nonumber\\
&&\left. \left. \frac{ \cos (\theta_{q_{1}} + \theta_{q_{n}}) \cos (\theta_{q_{n}} + \theta_{q_{n-1}}) \cdots \cos (\theta_{q_{j+2}} + \theta_{q_{j+1}})}{(k_0^\prime - 2\Lambda _{q_1} + i\varepsilon)(k_0^\prime - 2\Lambda _{q_n} + i\varepsilon) \cdots (k_0^\prime - 2\Lambda _{q_{j+2}} + i\varepsilon)(k_0^\prime -2 \Lambda _{q_{j+1}} + i\varepsilon)} \right ] \right \} . 
\label{n_ord_cumulant_convolution}
\end{eqnarray}

Due to the length of the calculations involved and the cumbersomeness of the final expression (\ref{n_ord_cumulant_convolution}), a short explanation is needed. 

The first factor under the integrals in the two square brackets is the WT of the projector, in accordance with the convolution rule (\ref{WT_convolution_product}). The inverse WT of the WT of the string of convolution products is taken in each of the two square brackets containing one of them. In front of the expression are the remaining greatest time and least time integrals. 

In the term of a sum with index $j=n$, in place of $\sin (\theta_{q_{n}} + \theta_{q_{n+1}}) $, there is the $\sin (\theta_{q_{n}} + \theta_{q_{1}}) $ factor. Furthermore, in the same term, under the integrals inside the second square brackets, there is only $1/(k_0^\prime - 2\Lambda _{q_1} + i\varepsilon)$ behind the WT of the projector. Similarly, in the $j=2$ term, under the integrals inside the first square brackets, there is only $1/(p_0^\prime - 2\Lambda _{q_2} + i\varepsilon)$ behind the WT of the projector. 

Factor $2^{n-1} n!$ in front of the sum is the number of ``bubble'' diagrams contained in the same term in the sum over $j$. Factor $i^n$ comes from factor $i$ in the WTs (\ref{WT_Green_2point_retarded}) of retarded two-point functions. 

Equation (\ref{n_ord_cumulant_convolution}) was obtained in our previous work \cite{Kuic1}, with the difference that there, local magnetic field perturbation (\ref{V_spins}) was introduced at the $N$-th spin site. Here, we have removed this constraint, which is ephemeral anyway due to translational symmetry, and we introduced the same perturbation at an arbitrary site $j$. It is not surprising that the same result was obtained as in \cite{Kuic1}. Following the same further calculations presented in \cite{Kuic1}, one can replicate all the results for a quench protocol with a perturbation switched on suddenly at $t=0$. We have rewritten (\ref{n_ord_cumulant_convolution}) here in order to highlight the difference in the analogous expressions for a perturbation that is switched on non-suddenly at $t=0$.

\section{Non-Sudden Quenches} \label{non-sudden_quenches}

Now we consider the case where the strength of the perturbation, local magnetic field $\delta h$, is switched on non-suddenly from $\delta h=0$ at $t=0$ to the value it has at time $t$. Let us assume that it is described by a time function $\delta h (t)$ with a Fourier transform $f(p)$ and an inverse Fourier transform as follows:
\begin{equation}
\delta h (t) = \frac{1}{2\pi}\int_{-\infty}^{\infty}f(p)e^{-ip t} dp . \label{delta h_Fourier_transform}
\end{equation}

Furthermore, assume that the Fourier transform $f(p)$ has the following properties:

\begin{itemize}
\item[(1)] \label{asumption_1} $f(p)$ is analytic everywhere in the lower complex semiplane except at the simple pole at $p = 0$ and possibly other simple poles with $\Im (p) < 0$, i.e., off the real axis.
\item[(2)] \label{asumption_2} $f(p)/p$ converges to zero at infinity in the lower complex semiplane (i.e., as $\left | p\right| \rightarrow \infty$ in the lower semiplane).
\end{itemize}

We immediately notice that if the function $f(p)$ has only one simple pole, located at $p=0$, one ends up with the Fourier transform $f(p) =i / (p + i\varepsilon )$ of the Heaviside unit step function $\theta (t)$,
\begin{equation}
\theta (t) = \frac{1}{2\pi}\int_{-\infty}^{\infty}\frac{i e^{-ip t}}{p + i\varepsilon } dp . \label{delta h_Fourier_transform_theta}
\end{equation}

Infinitesimally small $\varepsilon$ is added in the denominator of $f(p) =i / (p + i\varepsilon )$ to ensure that $\theta (t)$ has the correct form of a retarded function, and the $\varepsilon \rightarrow 0$ limit is taken after \mbox{the integration.}

The other important prototypical function of this class is 
\begin{equation}
\delta h (t) = \delta h \tanh(\eta t) . \label{delta h_tanh}
\end{equation}

This function is equal to zero $\delta h (0)=0$ at $t=0$, and it satisfies (1) and (2) in the lower (and also in the upper) complex semiplane. Its form for $t<0$ is not important since we are considering the time evolution starting from the initial state at $t=0$. The promptness of switching on the perturbation is controlled by the parameter $\eta$; sudden switching on corresponds to the $\eta \rightarrow \infty$ limit, which is  infinitely slow to the $\eta \rightarrow 0$ limit.

The Fourier transform of the function (\ref{delta h_tanh}) is 
\begin{equation}
f(p) = \frac{2\pi i \delta h}{\eta}\frac{1}{e^{\frac{\pi}{2\eta}p}-e^{-\frac{\pi}{2\eta}p}} . \label{delta h_ tanh_Fourier_transform}
\end{equation}

The function (\ref{delta h_ tanh_Fourier_transform}) is analytic everywhere except at the set of points $\{i 2\eta r : r \in \mathbb {Z}\}$ along the imaginary axis where it has simple poles.

Now consider a convolution product or $n$ projected retarded functions (\ref{Green_2point_retarded_projected}), but with a function $\delta h (t) $ inserted at all intermediate times $t_2, \ldots , t_{n}$,
\begin{equation}
C_t (t_1 - t_{n+1}) = \int_{-\infty}^{\infty} dt_2 \dots \int_{-\infty}^{\infty} dt_{n}\left [\prod_{l=1}^{n-1} G_{t, R, q_l} (t_l-t_{l+1}) \delta h (t_{l+1})\right ] G_{t, R, q_n} (t_{n}-t_{n+1}) . \label{convolution_product_Green_2point_retarded_projected_nonsudden}
\end{equation}

As deduced from reference \cite{Dadic1}, where expression (\ref{WT_convolution_product_gen}) is derived, the WT of (\ref{convolution_product_Green_2point_retarded_projected_nonsudden}) is
\begin{eqnarray}
&& C_t (p_0) = \int_{-\infty}^{\infty} \left (\prod_{j=1}^{n} dp_{0,j} \right ) \frac{\theta(t)}{\pi}\frac{\sin\left [2t \left (p_0 - \frac{p_{0,1} + p_{0,n}}{2} \right )\right ]}{p_0 - \frac{p_{0,1} + p_{0,n}}{2} } \cr\nonumber \\
&& \times \prod_{j=1}^{n-1}\left [G_{R, q_j}(p_{0,j}) \frac{1}{(2\pi)^2}\int _{-\infty}^{\infty}dp_{j+1}\frac{i}{p_{0,j} - p_{0, j +1} - p_{j +1}+ i \varepsilon} f(p_{j+1})\right ] \cr\nonumber\\
&& \times e^{-it(p_{0,1} - p_{0,n} + i(n-1)\varepsilon)} G_{R, q_n}(p_{0,n}) . \label{WT_convolution_product_Green_2point_retarded_projected_nonsudden}
\end{eqnarray}

We proceed to first take the two rightmost integrals over $p_{n}$ and $p_{0,n}$, and in that order, assuming (but without losing generality in the final conclusions) that $f(p)$ is given by (\ref{delta h_ tanh_Fourier_transform}). By closing the $p_{n}$ integration contour in the lower semiplane, and $p_{0,n}$ contour in the upper semiplane, we obtain
\begin{eqnarray}
&& C_t (p_0) = \delta h \sum_{r_{n-1}=0}^{\infty} \int_{-\infty}^{\infty} \left (\prod_{j=1}^{n-1} dp_{0,j} \right ) \frac{\theta(t)}{\pi}\frac{\sin\left [2t \left (p_0 - \frac{p_{0,1} + p_{0,n-1}+i2\eta r_{n-1} }{2} \right )\right ]}{p_0 - \frac{p_{0,1} + p_{0,n-1} +i2\eta r_{n-1} }{2} } \cr\nonumber \\
&& \times \prod_{j=1}^{n-2}\left [G_{R, q_j}(p_{0,j})  \frac{1}{(2\pi)^2}\int _{-\infty}^{\infty}dp_{j+1}\frac{i}{p_{0,j} - p_{0, j +1} -p_{j+1} + i \varepsilon} f(p_{j+1})\right ]  \cr\nonumber\\
&& \times e^{-it(p_{0,1} - p_{0,n-1} -i2\eta r_{n-1} + i(n-2)\varepsilon)}G_{R, q_{n-1}}(p_{0,n-1})  \cr\nonumber\\
&& \times (-1)^{r_{n-1}} 2^{\theta (r_{n-1} - \frac{1}{2})} G_{R, q_{n}}(p_{0,n-1}+i2\eta r_{n-1} + i \varepsilon) .  \label{WT_convolution_product_Green_2point_retarded_projected_nonsudden_1_int}
\end{eqnarray}

With an order of integrations and the closure of the contours as described above, if the contour of $p_{n}$ integration is closed in the lower semiplane, one avoids changing the position of poles of $i/(p_{0,n-1} - p_{0, n} -p_{n} + i \varepsilon)$ from the upper to a lower semiplane. Furthermore, the choice of the $p_n$ integration contour in the lower semiplane must be followed in an inverse Fourier transform (\ref{delta h_Fourier_transform}) for all $t >0$. Because of the factor $e^{-it(p_{0,1} - p_{0,n})}$ in the (\ref{WT_convolution_product_Green_2point_retarded_projected_nonsudden}), $p_{0,n}$ integration contour must be closed in the upper semiplane.  This also avoids closing the pole of the retarded function  $G_{R, q}(p_{0,n})$ by a $p_{0,n}$ integration contour. Eventually, one easily sees that the mutual order of $p_{n}$ and $p_{0,n}$ integrations is not important, but their integration contours must be chosen in the described way.

Heaviside unit step function  $\theta (r_{n-1} -\frac{1}{2}) $ in (\ref{WT_convolution_product_Green_2point_retarded_projected_nonsudden_1_int}) appears in (\ref{WT_convolution_product_Green_2point_retarded_projected_nonsudden_1_int}) because the $p_n$ contour must avoid going directly over the simple pole of $f(p_n)$ in (\ref{delta h_ tanh_Fourier_transform}) at $p=0$ at the real axis. This $p_n$ integral is a principle value integral and the contribution of the simple pole at $p_n=0$ is half of that of the residues at other poles, which are proportional to $(-1)^{r_{n-1}} 2^{\theta (r_{n-1} - \frac{1}{2})}$, hence the  $2^{\theta (r_{n-1} - \frac{1}{2})}$ factor.

After all the remaining integrations in (\ref{WT_convolution_product_Green_2point_retarded_projected_nonsudden_1_int}) are performed in the same way, we end \mbox{up with}

\vspace{-12pt}
\begin{eqnarray}
&& C_t (p_0) = (\delta h)^{n-1} \sum_{r_1, \ldots, r_{n-1} = 0}^{\infty} \int_{-\infty}^{\infty} dp_{0,1} \frac{\theta(t)}{\pi}\frac{\sin\left [2t \left (p_0 - p_{0,1} + i \eta (r_1 + \ldots + r_{n-1}) \right )\right ]}{p_0 - p_{0,1} + i \eta (r_1 + \ldots + r_{n-1}) }  \cr\nonumber \\
&& \times e^{-2\eta (r_1 + \ldots + r_{n-1} ) t}G_{R, q_1}(p_{0,1}) (-1)^{r_1} 2^{\theta (r_1 - \frac{1}{2})} G_{R, q_2}(p_{0,1}+ i2\eta r_1 + i \varepsilon) \cdots \cr\nonumber \\
&& \times (-1)^{r_{n-1}} 2^{\theta (r_{n-1} - \frac{1}{2})} G_{R, q_n}(p_{0,1}+i2\eta (r_1 + \ldots + r_{n-1}) + i (n-1)\varepsilon) . \label{WT_convolution_product_Green_2point_retarded_projected_nonsudden_n_int}
\end{eqnarray}

It is important to emphasize that all retarded two-point function WTs $G_{R, q}(p_{0})$ remain retarded after integrations that lead to (\ref{WT_convolution_product_Green_2point_retarded_projected_nonsudden_n_int}), albeit the decaying ones also appear (those acquiring nonzero width, i.e., a finite decay time) along with the non-decaying ones in the sums in (\ref{WT_convolution_product_Green_2point_retarded_projected_nonsudden_n_int}). 

However, tge contribution of the decaying two-point functions vanishes in exponentially short time because of the factor $e^{-2\eta (r_1 + \ldots + r_{n-1} ) t}$ in the convolution product (\ref{WT_convolution_product_Green_2point_retarded_projected_nonsudden_n_int}). Decay times of these contributions are proportional to $1/\eta$, which means that they are inversely proportional to the parameter $\eta$ controlling the rapidity of switching on the perturbation. The faster the perturbation is switched on ($\eta$ is greater) the shorter the decay times are.

Most importantly, we see that for non-sudden quenches, the same type of convolution product (\ref{WT_convolution_product}) of retarded projected functions (\ref{Green_2point_retarded_projected}), as in the case of a sudden quench, emerges in an exponentially short time scale. Furthermore, these conclusions do not depend on the exact form of the time-dependent perturbation strength (\ref{delta h_Fourier_transform}) as long as it satisfies properties (1) and (2). 

Notice also the important role of requirement (1); if there were other simple poles on the real axis except the one at $p=0$, the above would not be the case, and the $t \rightarrow \infty$ limit of (\ref{WT_convolution_product_Green_2point_retarded_projected_nonsudden_n_int}) would differ from (\ref{WT_convolution_product}). Some of $G_{R, q}(p_{0})$ in (\ref{WT_convolution_product_Green_2point_retarded_projected_nonsudden_n_int}) would acquire energy shifts and zero widths, and their contribution would not vanish in the $t \rightarrow \infty$ limit. Energy shifts are not prohibited by requirement (1); however, as long as simple poles that lead to them are moved off the real axis towards the lower semiplane, this in effect turns their contributions into exponentially decaying ones.

Of course, one has to take into account that, for non-sudden quenches, integrals~(\ref{delta h_Fourier_transform}) appear also at the greatest and lowest time vertices in (\ref{n_ord_cumulant_convolution}). This means that, in the limit $t \rightarrow \infty$, one obtains a somewhat modified expression with respect to (\ref{n_ord_cumulant_convolution}), due to the appearance of the integrals (\ref{delta h_Fourier_transform}) at vertices with the greatest and lowest time. In the case of the perturbation strength given by (\ref{delta h_tanh}), one gets, 
\begin{eqnarray}
&& \frac{(-i)^n}{n!} \int_0^\infty dt_1 \dots \int_0^\infty dt_n \bra{g_0^{\pm}}T[V_I(t_1) \dots V_I(t_n)]\ket{g_0^{\pm}}_C \cr \nonumber \\
&& = \frac{(-i)^n}{n!} \int_{0}^\infty dt_1 \int_{0}^\infty dt_n \theta(t_1 - t_n) \sum_{r_1,r_n = 0}^{\infty} (-1)^{r_1 +r_n} 2^{\theta (r_1 -\frac{1}{2}) + \theta (r_{n} -\frac{1}{2})} e^{-2\eta r_1 t_1 - 2\eta r_{n} t_n} \cr\nonumber\\
&& \left (- \frac{2^{n-1}n!(\delta h )^n i^n}{N^n}\right) \sum_{j=2}^{\mathrm{n}} \sum_{q_1, q_2 , \dots, q_n \in \Gamma ^{\pm }} \cr \nonumber\\ 
&& \left \{\sin (\theta_{q_1} + \theta_{q_2}) \left [ \frac{1}{2\pi}\int_{-\infty}^{\infty} dp_0 e^{-ip_o(t_1 - t_n)} \int_{-\infty}^{\infty} dp_0^\prime \frac{\theta(t)}{\pi}\frac{\sin(2t(p_0 - p_0^\prime))}{p_0 - p_0^\prime}\right . \right . \cr \nonumber\\
&& \left. \left . \frac{\cos (\theta_{q_2} + \theta_{q_3})\cos (\theta_{q_3} + \theta_{q_4}) \cdots \cos (\theta_{q_{j-1}} + \theta_{q_{j}})}{(p_0^\prime - 2\Lambda _{q_2} + i\varepsilon) (p_0^\prime - 2\Lambda _{q_3} + i\varepsilon) \cdots (p_0^\prime - 2\Lambda _{q_{j-1}} + i\varepsilon)(p_0^\prime -2 \Lambda _{q_j} + i\varepsilon)} \right ] \right . \cr \nonumber\\
&& \left . \sin (\theta_{q_{j}} + \theta_{q_{j+1}}) \left [ \frac{1}{2\pi}\int_{-\infty}^{\infty} dk_0 e^{-ik_o(t_1 - t_n)} \int_{-\infty}^{\infty} dk_0^\prime \frac{\theta(t)}{\pi}\frac{\sin(2t(k_0 - k_0^\prime))}{k_0 - k_0^\prime} \right . \right . \cr \nonumber\\
&&\left. \left. \frac{ \cos (\theta_{q_{1}} + \theta_{q_{n}}) \cos (\theta_{q_{n}} + \theta_{q_{n-1}}) \cdots \cos (\theta_{q_{j+2}} + \theta_{q_{j+1}})}{(k_0^\prime - 2\Lambda _{q_1} + i\varepsilon)(k_0^\prime - 2\Lambda _{q_n} + i\varepsilon) \cdots (k_0^\prime - 2\Lambda _{q_{j+2}} + i\varepsilon)(k_0^\prime -2 \Lambda _{q_{j+1}} + i\varepsilon)} \right ] \right \} . 
\label{n_ord_cumulant_convolution_infinite time} 
\end{eqnarray}

From this point on one can proceed with the application of the same calculational and resummation procedure developed in \cite{Kuic1}. We only briefly describe it, since it is not necessary to repeat the calculations here, for all the details see \cite{Kuic1}. 

One first calculates the integrals over $p_0$ (or over $p_0^\prime$) in the first square bracket, and the integrals over $k_0$ (or over $k_0^\prime$) in the second square bracket of (\ref{n_ord_cumulant_convolution_infinite time}). Then, one proceeds to resum the perturbative expansion in the exponent of (\ref{G_cumulant_exp}) by resumming the generalized Schwinger--Dyson equations for the two-point retarded functions $G_{R, q}(p_{0})$ with intermediate vertex insertions in the ``bubble'' diagrams, thus resumming over all ``bubble'' diagrams appearing at all orders of the perturbative expansion (\ref{G_cumulant_exp}). Finally, under the analyticity assumptions formulated in \cite{Kuic1}, one calculates the remaining $p_0$ and $k_0$ integrals, and then the $t_1$ and $t_n$ time integrals. 

To briefly summarize, by first taking into account that in the long-time limit $t >> 1/\eta$ all exponentially decaying contributions vanish, and by following the described procedure, one arrives at the expression

\vspace{-12pt}
\begin{eqnarray}
&& \log[\mathcal{G}(t)] = i \frac{\delta h}{N}\left ( \sum_{q \in \Gamma \pm } \cos 2\theta _q \right ) t - \frac{2 (\delta h)^2}{N^2} \sum_{q_1, q_2, q_3, q_4 \in \Gamma ^{\pm }} \left \{\sin (\theta_{q_1} + \theta_{q_2}) \sin (\theta_{q_{3}} + \theta_{q_{4}}) \right . \cr \nonumber\\ 
&& \left. \times \left [ \left (R_{q_2, q_3}({\bf \hat {A}}(\delta h , N)) + \sum_{n=2}^{\infty} R_{q_2; q_2, q_3}({\bf A}( 2\Lambda _{q_2}, \delta h , N))_{(n-1) \ \textrm{amp.}} \frac{1}{(n-1)!}\frac{d^{n-1}}{(d 2\Lambda _{q_2})^{n-1}} \right) \right. \right . \cr \nonumber \\ 
&& \left. \left . \times \left (R_{q_1, q_4}({\bf \hat {A}}(\delta h , N)) + \sum_{m=2}^{\infty}R_{q_1; q_1, q_4}({\bf A}( 2\Lambda _{q_1}, \delta h , N))_{(m-1) \ \textrm{amp.}} \frac{1}{(m-1)!}\frac{d^{m-1}}{(d 2\Lambda _{q_1})^{m-1}} \right ) \right . \right . \cr \nonumber\\
&& \left . \left . \times \mathcal{F}_{\infty} ( 2\Lambda _{q_1}, 2\Lambda _{q_2}, t) \right . \right. \cr\nonumber \\
&& \left . \left . + 2 \left (R_{q_2, q_3}({\bf \hat {A}}(\delta h , N)) + \sum_{n=2}^{\infty} R_{q_2; q_2, q_3}({\bf A}( 2\Lambda _{q_2}, \delta h , N))_{(n-1) \ \textrm{amp.}} \frac{1}{(n-1)!}\frac{d^{n-1}}{(d 2\Lambda _{q_2})^{n-1}} \right) \right . \right . \cr \nonumber \\
&& \left . \left . \times \left ( \sum_{q_6 \in \Gamma ^{\pm }} ^{\Lambda _{q_6} \ne \Lambda _{q_1} } \sum_{m=1}^{\infty}R_{q_1; q_6, q_4}({\bf A}( 2\Lambda _{q_6}, \delta h , N))_{m \ \textrm{amp.}} \frac{1}{(m-1)!}\frac{d^{m-1}}{(d 2\Lambda _{q_6})^{m-1}} \right ) \frac{\mathcal{F}_{\infty} ( 2\Lambda _{q_6}, 2\Lambda _{q_2}, t) }{ 2\Lambda _{q_6} - 2\Lambda _{q_1}} \right . \right . \cr \nonumber \\
&& \left . \left . + \left ( \sum_{q_5, q_6 \in \Gamma ^{\pm }} ^{\Lambda _{q_5} \ne \Lambda _{q_2}, \Lambda _{q_6} \ne \Lambda _{q_1} } \sum_{n=1}^{\infty}R_{q_2; q_5, q_3}({\bf A}( 2\Lambda _{q_5}, \delta h , N))_{n \ \textrm{amp.}} \frac{1}{(n-1)!}\frac{d^{n-1}}{(d 2\Lambda _{q_5})^{n-1}} \right . \right. \right. \cr \nonumber\\
&& \left. \left . \left . \times \sum_{m=1}^{\infty}R_{q_1; q_6, q_4}({\bf A}( 2\Lambda _{q_6}, \delta h , N))_{m \ \textrm{amp.}} \frac{1}{(m-1)!}\frac{d^{m-1}}{(d 2\Lambda _{q_6})^{m-1}}\right ) \right . \right . \cr\nonumber \\
&& \left. \left . \times \frac{\mathcal{F}_{\infty} ( 2\Lambda _{q_6}, 2\Lambda _{q_5}, t) }{ (2\Lambda _{q_6} - 2\Lambda _{q_1})(2\Lambda _{q_5} - 2\Lambda _{q_2})} \right ] \right \} . \label{G_cumulant_exp_resummed_int_total_infinite time} 
\end{eqnarray}

Let us briefly explain the symbols in (\ref{G_cumulant_exp_resummed_int_total_infinite time}). A more detailed explanation is given in \cite{Kuic1}. 
$R_{q, q^\prime}(\mathbf {\hat{A}} ( \delta h , N))$ are the matrix elements of $\mathbf {R}(\mathbf {\hat{A}} ( \delta h , N))$, the sum of infinite geometric series of the matrix $\mathbf {\hat{A}} ( \delta h , N)$. $\mathbf {R}(\mathbf {\hat{A}} ( \delta h , N))$ is obtained after the resummation of the generalized Schwinger--Dyson equations for the two-point retarded functions $G_{R, q}(p_{0})$, and subsequently, $p_0$ and $k_0$ integrations. 

$\mathbf {\hat{A}} ( \delta h , N))$ is the $N \times N$ matrix with the elements
\begin{equation}
\hat{A}_{q_1, q_2} (\delta h , N) = \left \{ \begin{array}{l@{\,,\qquad }l} \frac{\frac{2 \delta h}{N} \cos (\theta_{q_1} + \theta_{q_2})} {( 2\Lambda _{q_1} - 2\Lambda _{q_2} + i\varepsilon)} & q_1, q_2 \in \Gamma ^{\pm }, 2\Lambda _{q_1} \neq 2\Lambda _{q_2} \\
0 & q_1, q_2 \in \Gamma ^{\pm }, 2\Lambda _{q_1} = 2\Lambda _{q_2} \end{array} \right . . \label{mat_A_q}
\end{equation}

$R_{q_1; q_2, q_3}({\bf A}( 2\Lambda _{q_2}, \delta h , N))_{n \ \textrm{amp.}}$ is a part of the matrix element of the geometric series $R_{q_1, q_3}({\bf A}(p_0, \delta h , N))$ containing only $n$-th order singular terms with ``amputated'' legs, which then renders them nonsingular, i.e.,
\begin{equation}
R_{q_1; q_2, q_3}({\bf A}( 2\Lambda _{q_2}, \delta h , N))_{n \ \textrm{amp.}} = \lim_{p_0 \rightarrow 2\Lambda _{q_2} }\left [(p_0 - 2\Lambda _{q_2} )^n R_{q_1, q_3}({\bf A}(p_0, \delta h , N))_{n \ \textrm{sing.}} \right ] . \label{AppB_Res4}
\end{equation}

${\bf A} (p_0, \delta h , N)$ is the matrix with the elements
\begin{equation}
A_{q_1, q_2} (p_0, \delta h , N) = \frac{\frac{2 \delta h}{N} \cos (\theta_{q_1} + \theta_{q_2})} {(p_0 - 2\Lambda _{q_2} + i\varepsilon)} , \qquad q_1, q_2 \in \Gamma ^{\pm } . \label{mat_A} 
\end{equation}

Function $ \mathcal{F} (x, y, t) $ is a result of $t_1$ and $t_n$ integrations. It is given by
\begin{eqnarray}
&& \mathcal{F} (x, y, t) = \sum_{r_1,r_n = 0}^{\infty} (-1)^{r_1 + r_{n}} 2^{\theta (r_1 -\frac{1}{2})+ \theta (r_{n} -\frac{1}{2})} \cr\nonumber\\
&& \int_{0}^t dt_1 \int_{0}^t dt_n \theta(t_1 - t_n) e^{-i(x+y)(t_1-t_n)}e^{-2\eta r_1 t_1 - 2\eta r_{n} t_n} \cr\nonumber\\
&& = \sum_{r_1,r_n = 0}^{\infty} (-1)^{r_1 + r_{n}} 2^{\theta (r_1 -\frac{1}{2})+ \theta (r_{n} -\frac{1}{2})} \cr\nonumber\\ 
&& \left [ \frac{1-e^{- 2\eta r_1 t} e^{-i(x + y) t }}{(x+y - i2\eta r_1)(x+y + i2\eta r_n)} + \frac{1 - e^{- 2\eta(r_1+r_n)t}}{(x+y + i 2\eta r_n)(i2 \eta (r_1 + r_n )) } \right ] . \label{fun_time_int_finite}
\end{eqnarray}

$ \mathcal{F}_{\infty} (x, y, t) $ in (\ref{G_cumulant_exp_resummed_int_total_infinite time}) is the long-time $t >> 1/\eta$ limit of (\ref{fun_time_int_finite}). It is obtained by removing all exponentially decaying terms from (\ref{fun_time_int_finite}),
\begin{eqnarray}
&& \mathcal{F}_{\infty} (x, y, t) = \frac{1- e^{-i(x + y) t }}{(x+y)^2} -\frac{it}{x+y} \cr\nonumber\\ 
&& + \sum_{r_n=1}^{\infty} (-1)^{r_n} 2^{\theta (r_n -\frac{1}{2})} \left [\frac{1}{i2\eta r_n (x+y)} - \frac{e^{-i(x + y) t }}{(x+y)(x+y + i2\eta r_n)} \right ] \cr\nonumber\\ 
&& + \sum_{r_1=1, r_n = 0}^{\infty} (-1)^{r_1 + r_n} 2^{\theta (r_1 -\frac{1}{2}) + \theta (r_{n} -\frac{1}{2}) } \frac{1}{(x+y - i 2\eta r_1)(i2 \eta (r_1 + r_n )) } . \label{fun_time_int_infinite}
\end{eqnarray}

Furthermore, finally, derivatives in (\ref{G_cumulant_exp_resummed_int_total_infinite time}) are over all the variables $x$ and $y$ of the function $\mathcal{F}_{\infty} (x, y, t)$ and other functions that appear to the right of the derivative symbols. 

It is important to make some further observations about (\ref{G_cumulant_exp_resummed_int_total_infinite time}) and (\ref{fun_time_int_infinite}). Part of (\ref{fun_time_int_infinite}) consisting of the first two terms is equal to the function appearing in place of $ \mathcal{F}_{\infty} (x, y, t) $ in the similar expression of $\log[\mathcal{G}(t)]$ for the sudden quench obtained in \cite{Kuic1}. Otherwise, the sudden quench expression for $\log[\mathcal{G}(t)]$, and the expression (\ref{G_cumulant_exp_resummed_int_total_infinite time}) of the same quantity for non-sudden quenches in the long-time limit, are completely analogous. The only difference comes from the remaining terms of (\ref{fun_time_int_infinite}) after the first two.

These remaining terms appear as the contribution of simple poles of the perturbation strength Fourier transform (\ref{delta h_ tanh_Fourier_transform}) in the lower semiplane, additional to that of the simple pole at $p=0$, which is strictly responsible for the first two terms in (\ref{fun_time_int_infinite}). Among these remaining terms, the only time-dependent term is the one proportional to $e^{-i(x+y)t}$, which comes from the vertex with the greatest time in ``bubble'' diagrams. 

The same is true for all other functions (\ref{delta h_Fourier_transform}) satisfying properties (1) and (2). In other words, as long as the Fourier transform $f(p)$ satisfies (1) and (2), the first two terms in (\ref{fun_time_int_infinite}) do not depend on the exact form of $\delta h (t)$. All other terms depend on it through the contribution of the lower semiplane simple poles of $f(p)$, i.e., simple poles with $\Im (p) < 0$. 

A specific form of $ \mathcal{F}_{\infty} (x, y, t) $ for each function $\delta h (t)$ of the whole class of functions satisfying (1) and (2) is easily obtainable from (\ref{fun_time_int_infinite}). In the sums over the contributions of the simple poles of $ (\ref{delta h_ tanh_Fourier_transform})$ in (\ref{fun_time_int_infinite}), just replace $-i2\eta r$ (for all $ r \geq 1$) with the positions of all $\Im (p) < 0$ lower semiplane simple poles of $f(p)$. Furthermore, replace the factors $(-1)^r 2^{\theta (r - \frac{1}{2})}$ with contributions of residues of $f(p)$ at these poles.

So, up to these specific changes of the particular form of $ \mathcal{F}_{\infty} (x, y, t) $, the behavior of $\log[\mathcal{G}(t)]$ is in all other respects universally described in the long-time limit by expression~(\ref{G_cumulant_exp_resummed_int_total_infinite time}). This is true for a class of perturbation strength switching on functions $\delta h (t)$ that satisfy (1) and (2) and describe non-sudden quenches starting at $t=0$. 

This class includes the prototypical function $\delta h (t) = \delta h \tanh(\eta t)$, with the Fourier transform given by (\ref{delta h_ tanh_Fourier_transform}). It also includes its respective sudden quench $\eta \rightarrow \infty$ limit, function $\delta h (t) = \delta h \theta (t)$, with the Fourier transform of $\theta (t)$ given by (\ref{delta h_Fourier_transform_theta}). 

Sudden quench expression for $\log[\mathcal{G}(t)]$ are completely analogous to (\ref{G_cumulant_exp_resummed_int_total_infinite time}), with the only difference that the function (\ref{fun_time_int_infinite}) is replaced only by its first two terms. That expression was previously obtained in \cite{Kuic1}, it is valid at any time during the sudden quench and not restricted only to a long-time limit.

\section{Disorder and Global Quenches} \label{disorder_quenches}

Previously we have considered only local quenches, where perturbation is switched on by a sudden or non-sudden local change of the magnetic field $\delta h$ only at a single spin site $j$. With a slight and, as we shall see, rather straightforward modification of the calculational procedure leading to (\ref{G_cumulant_exp_resummed_int_total_infinite time}), we can extend the developed FTPFT perturbative and resummation approach to include disorder-type quenches, and eventually, global ones. 

Perturbation (\ref{V_spins}) is modified to describe disorder type perturbation over the \mbox{entire chain,}
\begin{equation}
V= \delta h (t) \sum_{j=1}^N \xi_j \sigma_j^z , \label{V_spins_disorder}
\end{equation}

Disorder variables $\{\xi_j : j=1, \dots, N \}$ are randomly distributed in the interval $[-1,1]$, and the perturbation strength parameter $ \delta h (t)$ controls the switching on the disorder perturbation (\ref{V_spins_disorder}).

In terms of Bogoliubov fermion operators that diagonalize the Hamiltonian (\ref{XY_Hamiltonian}), perturbation (\ref{V_spins_disorder}) reads
\begin{equation}
V = \frac{\delta h}{N} \sum_{q, q^\prime \in \Gamma ^\pm }\chi(q-q^{\prime}) \left( \begin{array}{cc} \eta _{q}^\dagger & \eta _{-q} \end{array} \right ) a_{q, q^\prime} \left( \begin{array}{c} \eta _{q^\prime} \\ \eta _{-q^\prime}^\dagger \end{array} \right ) , \label{V_Bogoliubov_disorder}
\end{equation}
where $\chi(q-q^{\prime})$ is the discrete Fourier transform of $\xi _j$, 
\begin{equation}
\chi(q-q^{\prime}) = \sum_{j=1}^N \xi_j e^{i(q-q^{\prime})j} . \label{chi_random}
\end{equation}
$\chi(q-q^{\prime})$ are thus themselves (complex) random variables on $N \times N$ lattice $\Gamma ^\pm \times \Gamma ^\pm $.

As deduced from (\ref{V_Bogoliubov_disorder}) and the subsequent arguments presented in Sections \ref{perturbative_calculations}--\ref{non-sudden_quenches}, all previous results for sudden and non-sudden local quenches are easily modified to include the disorder-type quenches by a simple replacement of $\delta h e^{i(q-q^{\prime})j}$ with $\delta h \chi(q-q^{\prime})$ in the vertex factors $ \pm i (\delta h /N)e^{i(q-q^{\prime})j} \sin (\theta _q + \theta _{q^\prime})$ and $ \pm (\delta h /N)e^{i(q-q^{\prime})j} \cos (\theta _q + \theta _{q^\prime})$. As result of this, $\delta h $ is then replaced with $\delta h \chi(q-q^{\prime})$ in all expressions starting from (\ref{n_ord_cumulant_convolution}).

The variables $\chi(q-q^{\prime})$ in (\ref{chi_random}) can be represented as the elements of an $N \times N$ Hermitian matrix, since $\chi(q-q^{\prime}) = \chi^\ast (q^{\prime} - q)$. Matrix (\ref{mat_A_q}), with $\delta h $ replaced in its elements by $\delta h \chi(q-q^{\prime})$, remains skew-Hermitian, and therefore, unitarily diagonalizable. This last fact enables the implementation of analyticity assumptions formulated in \cite{Kuic1} to all integration and resummation procedures. This leads to the same final expression for $\log[\mathcal{G}(t)]$ for disorder-type quenches as (\ref{G_cumulant_exp_resummed_int_total_infinite time}), where the only difference is the substitution of $\delta h $ with $\delta h \chi(q-q^{\prime})$. 

For non-sudden disorder-type quenches, the behavior of LE universal with a sudden type of the same disorder quench emerges in the long-time limit regardless of the perturbation strength function $\delta h(t)$. This happens as long as $\delta h(t)$ satisfies properties (1) and (2), in the same way as described in Section \ref{non-sudden_quenches} for local quenches.
It is important to emphasize that all this applies to global quenches also. They are considered a special case of quenches described in this section as follows: perturbation given by (\ref{V_spins_disorder})--(\ref{chi_random}) includes global quenches as a limiting case when a particular set of values of variables $\xi_j$, e.g., $\{\xi_j = 1 : j=1, \dots, N \}$, \mbox{is certain.}

\section{Conclusions} \label{conclusions}

We will briefly summarize the most important results of this work and the conclusions derived from it. They are concerning the class of non-sudden local quench protocols described by the magnetic field perturbation strength functions $\delta h(t)$ that we discussed. The short-time behavior (short compared to the time scale of switching on the perturbation) of the LE of Ising and XY chains when quenching their nondegenerate ground states is dominated by excitations appearing in the ``bubble'' diagrams at all orders of the perturbation expansion. These excitations are propagated by retarded two-point fermion functions with shifted energies and decay rates that depend on the details of the functions $\delta h(t)$. Under the requirements (1) and (2) on perturbation strength functions $\delta h(t)$ formulated in Section \ref{non-sudden_quenches}, these excitations are a feature that depends only on the positions of the poles of Fourier transform of $\delta h(t)$ in the lower complex semiplane. Most importantly, these excitations all decay exponentially, with decay times inversely proportional to their perpendicular distance from the real axis. 

So, in the long-time limit, which means times much larger than decay times of these excitations which depend on the quench protocol, LE is purified from them, and what is left is essentially a sudden quench behavior of LE. To be more precise, it is not exactly the same as the sudden quench LE, it is modified, but in a non-substantial way, with the only time-dependent modification coming from the vertex with the greatest time of ``bubble'' diagrams. These modifications are systematically taken into account and are non-substantial, in the sense that they are subdominant and do not entail any kind of different time-dependent behavior, compared to the main part, which is the same as for the sudden quench LE obtained in \cite{Kuic1}. Such finite time effects are typical in FTPFT \cite{Dadic1,Dadic2,Dadic3,Dadic4}. They are, of course, absent in standard applications of quantum field theory, where perturbation (interactions) is switched on adiabatically starting from $t= - \infty$ \cite{Lancaster, Coleman, Le Bellac}.

Thus, for a class of quenches described by the perturbation strength functions $\delta h(t)$ satisfying quite general properties (1) and (2), the long-time behavior of the LE is common and universal with a sudden quench. Under the same conditions (1) and (2) on $\delta h(t)$, the analogous form of universal behavior is obtained for disorder-type quenches, and ultimately for global quenches.

In our understanding, the physical reason behind this long-time limit universal behavior of LE has at least three components. Firstly, finite lifetimes of the non-sudden quench excitations which appear in the ``bubble'' diagrams are a direct result of properties (1) and (2) of  the Fourier transform of perturbation strength functions $\delta h(t)$. To be specific, due to (1), with the exception of excitations common with the sudden quench that correspond to a simple pole at $p=0$, all other non-sudden quench excitations have finite lifetimes and decay in exponentially short time compared to the long-time limit of interest. Only in tge adiabatic limit of infinitely slow switching on the perturbation we expect them to become stable. In a certain sense this is similar to the adiabatic stability of quasiparticles encountered in the context of Landau Fermi liquid theory \cite{Lancaster, Coleman}, but only partially. For example, in the case of the prototypical function $\delta h (t) = \delta h \tanh(\eta t)$, there is an infinite variety of such excitations, and this is clearly at odds with adiabatic continuity, as there is no one-to-one correspondence between particles and quasiparticles present in Landau Fermi liquid theory. As a consequence, this universal behavior of LE is destroyed in the adiabatic limit $\eta \rightarrow 0^+$.

The second component that enables the long-time limit universal behavior of the LE is that, due to causality, only retarded two-point fermion functions appear in the ``bubble'' diagrams, and the only time-dependent modification with respect to the sudden quench case comes from the vertex with greatest time. This is a non-sudden quench generalization (\ref{WT_convolution_product_Green_2point_retarded_projected_nonsudden_n_int}) of the property (\ref{WT_convolution_product}) of WTs of the convolution products of projected retarded two-point functions, which obey the same rule as the functions themselves and are also retarded functions \cite{Dadic1,Dadic2,Dadic3,Dadic4}. The third component is the nature of transverse magnetic field perturbation of the spin chain, which can be described as an external field on which fermion excitations are scattered. These three components together allow the non-sudden quench convolution rule (\ref{WT_convolution_product_Green_2point_retarded_projected_nonsudden_n_int})  presented in this paper.

Having in mind the possibility of spin chains being used in similar concepts of quantum technology applications of energy storage \cite{Catalano1} (and/or information storage and transfer \cite{Sacco1}), the main finding of this work is that such concepts are probably much less sensitive on the details of operating protocols and on the degree of experimental control over them than is expected. Proposals like \cite{Rossini3}  for measuring the LE of Ising and XY chains by coupling them to an auxiliary two-level system, engineered with atoms in optical lattices \cite{Duan1}, makes such concepts of potential interest. This is even more interesting, since the complex amplitude of LE is equal to the characteristic function of the probability of the work done on the system \cite{Silva1}, and a protocol-independent  
universal behavior has already been found near the critical point of Ising chain \cite{Smacchia1}, at least for the low energy part of the distribution. Based around GGE and light-cone behavior, there are different arguments for the robustness of the form of response of fermionic systems on quench protocols near the critical points \cite{Porta2, Porta3}. Our results are perturbative but are obtained by a full resummation of the perturbative expansion, valid under simple and general analyticity assumptions formulated in our previous paper \cite{Kuic1}. Of course, more detailed numerical studies of this issue will be carried out in future work.

\vspace{6pt}

\acknowledgments{D.K. wishes to express particular thanks to Ivan Dadi\'{c}, Vanja Mari\'{c}, Gianpaolo Torre, Fabio Franchini and Salvatore Marco Giampaolo for discussions and motivation.} 



\begin{thebibliography}{999}
\bibitem{Polkovnikov1} Polkovnikov, A.; Sengupta, K.; Silva, A.; Vengalattore, M. Colloquium: Nonequilibrium dynamics of closed interacting quantum systems. \emph{Rev. Mod. Phys.} \textbf{2011}, \emph{83}, 863.
\bibitem{Mitra2} Mitra, A. Quantum Quench Dynamics. \emph{Annu. Rev. Condens. Matter Phys.} \textbf{2018}, \emph{9}, 245. 
\bibitem{Das1} Das, S. Quantum Quench and Universal Scaling. In \emph{Oxford Research Encyclopedia of Physics}; Oxford University Press: Oxford, UK,  2020 
\bibitem{Essler1} Essler, F.H.L.; Fagotti, M. Quench dynamics and relaxation in isolated integrable quantum spin chains. \emph{J. Stat. Mech.} \textbf{2016}, \emph{2016},~064002.
\bibitem{Mitra1} Mitra, A. Correlation functions in the prethermalized regime after a quantum quench of a spin chain. \emph{Phys. Rev. B} \textbf{2013}, \emph{87},~205109.
\bibitem{Marcuzzi1} Marcuzzi, M.; Marino, J.; Gambassi, A.; Silva, A. Prethermalization in a Nonintegrable Quantum Spin Chain after a Quench. \emph{Phys. Rev. Lett.} \textbf{2013}, \emph{111}, 197203.
\bibitem{Bertini1} Bertini, B.; Fagotti, M. Pre-relaxation in weakly interacting models. \emph{J. Stat. Mech.} \textbf{2015}, \emph{2015}, P07012.
\bibitem{Nadkishore1} Nandkishore, R.; Huse, D.A. Many-Body Localization and Thermalization in Quantum Statistical Mechanics. \emph{Annu. Rev. Condens. Matter Phys.} \textbf{2015}, \emph{6}, 15.
\bibitem{Yang1} Yang, Z.-C.; Hamma, A.; Giampaolo, S.M.; Mucciolo, E.R.; Chamon, C. Entanglement complexity in quantum many-body dynamics, thermalization, and localization. \emph{Phys. Rev. B} \textbf{2017}, \emph{96,} 020408.
\bibitem{Zunkovic1} \v{Z}unkovi\v{c}, B.; Silva, A.; Fabrizio, M. Dynamical phase transitions and Loschmidt echo in the infinite-range XY model. \emph{Philos. Trans. R. Soc. A} \textbf{2016}, \emph{374}, 20150160.
\bibitem{Jafari1} Jafari, R. Dynamical Quantum Phase Transition and Quasi Particle Excitation. \emph{Sci. Rep.} \textbf{2019}, \emph{9}, 2871.
\bibitem{Senegupta1} Sengupta, K.; Powell, S.; Sachdev, S. Quench dynamics across quantum critical points. \emph{Phys. Rev. A} \textbf{2004}, \emph{69}, 053616.
\bibitem{Guo1} Guo, H.L.; Liu, Z.; Fan, H.; Chen, S. Correlation properties of anisotropic XY model with a sudden quench. \emph{Eur. Phys. J. B} \textbf{2011}, \emph{79},~503. 
\bibitem{Foini1} Foini, L.; Cugliandolo, L.F.; Gambassi, A. Fluctuation-dissipation relations and critical quenches in the transverse field Ising chain. \emph{Phys. Rev. B} \textbf{2011}, \emph{84}, 212404.
\bibitem{Paul1} Paul, S.; Titum, P.; Maghrebi, M. Hidden quantum criticality and entanglement in quench dynamics. \emph{Phys. Rev. Res.} \textbf{2024}, \emph{6}, L032003.
\bibitem{Ding1} Ding, C. Dynamical quantum phase transition from a critical quantum quench. \emph{Phys. Rev. B} \textbf{2020}, \emph{102}, 060409.
\bibitem{Porta1} Porta, S.; Cavaliere, F.; Sasseti, M.; Ziani, N.T. Topological classification of dynamical quantum phase transitions in the xy chain. \emph{Sci. Rep.} \textbf{2020}, \emph{10}, 12766.
\bibitem{Lupo1} Lupo, C.; Schir\'o, M. Transient Loschmidt echo in quenched Ising chains. \emph{Phys. Rev. B} \textbf{2016}, \emph{94}, 014310. 
\bibitem{Silva1} Silva, A. The statistics of the work done on a quantum critical system by quenching a control parameter. \emph{Phys. Rev. Lett.} \textbf{2008}, \emph{101}, 120603.
\bibitem{Gambassi1} Gambassi, A.; Silva, A. Statistics of the Work in Quantum Quenches, Universality and the Critical Casimir Effect. \emph{arXiv} \textbf{2011}
, arXiv:1106.2671v1.
\bibitem{Smacchia1} Smacchia, P.; Silva, A. Work distribution and edge singularities for generic time-dependent protocols in extended systems. \emph{Phys. Rev. E} \textbf{2013}, \emph{88}, 042109.
\bibitem{groupLE} Torre, G.; Mari\'{c}, V.; Kui\'{c}, D.; Franchini, F.; Giampaolo, S.M. Odd thermodynamic limit for the Loschmidt echo. 
\emph{Phys. Rev. B} \textbf{2022}, \emph{105}, 184424.
\bibitem{Catalano1} Catalano, A.G.; Giampaolo, S.M.; Morsch, O.; Giovannetti, V.; Franchini, F. Frustrating Quantum Batteries. 
\emph{PRX Quantum} \textbf{2024}, \emph{5},~030319.
\bibitem{Fagotti1} Fagotti, M.; Calabrese, P. Evolution of entanglement entropy following a quantum quench: Analytic results for the XY chain in a transverse magnetic field. \emph{Phys. Rev. A} \textbf{2008}, \emph{78}, 010306.
\bibitem{Rossini1} Rossini, D.; Silva, A.; Mussardo, G.; Santoro, G. Effective thermal dynamics following a quantum quench in a spin chain. \emph{Phys. Rev. Lett.} \textbf{2009}, \emph{102}, 127204.
\bibitem{Rossini2} Rossini, D.; Suzuki, S.; Mussardo, G.; Santoro, G.E.; Silva, A. Long time dynamics following a quench in an integrable quantum spin chain: Local versus non-local operators and effective thermal behaviour. \emph{Phys. Rev. B} \textbf{2010}, \emph{82}, 144302.
\bibitem{Campos1} Venuti, L.C.; Zanardi, P. Unitary equilibrations: Probability distribution of the Loschmidt echo. \emph{Phys. Rev. A} \textbf{2010}, \emph{81}, 022113.

\pagebreak
\bibitem{Canovi1} Canovi, E.; Rossini, D.; Fazio, R.; Santoro, G.E.; Silva, A. Quantum Quenches, Thermalization and Many-Body Localization. \emph{Phys. Rev. B} \textbf{2011}, \emph{83}, 094431.
\bibitem{Campos3} Venuti, L.C.; Jacobson, N.T.; Santra, S.; Zanardi, P. Exact Infinite-Time Statistics of the Loschmidt Echo for a Quantum Quench. \emph{Phys. Rev. Lett.} \textbf{2011}, \emph{107}, 010403.
\bibitem{Calabrese1} Calabrese, P.; Essler, F.H.L.; Fagotti, M. Quantum Quench in the Transverse Field Ising Chain. \emph{Phys. Rev. Lett.} \textbf{2011}, \emph{106}, 227203.
\bibitem{Igloi1} Igl\'oi, F.; Rieger, H. Long-Range correlations in the nonequilibrium quantum relaxation of a spin chain. \emph{Phys. Rev. Lett.} \textbf{2000}, \emph{85},~3233.
\bibitem{Igloi2} Igl\'oi, F.; Rieger, H. Quantum relaxation after a quench in systems with boundaries. \emph{Phys. Rev. Lett.} \textbf{2011}, \emph{106}, 035701.
\bibitem{Riegler1} Rieger, H.; Igl\'oi, F. Semiclassical theory for quantum quenches in finite transverse Ising chains. \emph{Phys. Rev. B} \textbf{2011}, \emph{84}, 165117.
\bibitem{Schuricht1} Schuricht, D.; Essler, F.H.L. Dynamics in the Ising field theory after a quantum quench. \emph{J. Stat. Mech.} \textbf{2012}, \emph{2012}, 
P04017.
\bibitem{Calabrese2} Calabrese, P.; Essler, F.H.L.; Fagotti, M. Quantum quench in the transverse field Ising chain: I. Time evolution of order parameter correlators. \emph{J. Stat. Mech.} \textbf{2012}, \emph{2012}, P07016.
\bibitem{Calabrese3} Calabrese, P.; Essler, F.H.L.; Fagotti, M. Quantum quenches in the transverse field Ising chain: II. Stationary state properties. \emph{J. Stat. Mech.} \textbf{2012}, \emph{2012}, P07022.
\bibitem{Fagotti2} Fagotti, M. Finite-size corrections versus relaxation after a sudden quench. \emph{Phys. Rev. B} \textbf{2013}, \emph{87}, 165106.
\bibitem{Heyl1} Heyl, M.; Polkovnikov, A.; Kehrein, S. Dynamical Quantum Phase Transitions in the Transverse-Field Ising Model. \emph{Phys. Rev. Lett.} \textbf{2013}, \emph{110}, 135704.
\bibitem{Rossini3} Rossini, D.; Calarco, T.; Giovannetti, V.; Montangero, S.; Fazio, R. Decoherence induced by interacting quantum spin baths. \emph{Phys. Rev. A} \textbf{2007}, \emph{75}, 032333.
\bibitem{Goold1} Goold, J.; Fogarty, T.; Gullo, N.L.; Paternostro, M.; Busch, T. Orthogonality catastrophe as a consequence of qubit embedding in an ultracold Fermi gas. \emph{Phys. Rev. A} \textbf{2011}, \emph{84}, 063632.
\bibitem{Knap1} Knap, M.; Shashi, A.; Nishida, Y.; Imambekov, A.; Abanin, D.A.; Demler, E. Time-Dependent Impurity in Ultracold Fermions: Orthogonality Catastrophe and Beyond. \emph{Phys. Rev. X} \textbf{2012}, \emph{2}, 041020.
\bibitem{Knap2} Knap, M.; Kantian, A.; Giamarchi, T.; Bloch, I.; Lukin, M.D.; Demler, E. Probing Real-Space and Time-Resolved Correlation Functions with Many-Body Ramsey Interferometry. \emph{Phys. Rev. Lett.} \textbf{2013}, \emph{111}, 147205.
\bibitem{Dora1} D\'ora, B.; Pollmann, F.; Fort\'agh, J.; Zar\'and, G. Loschmidt Echo and the Many-Body Orthogonality Catastrophe in a Qubit-Coupled Luttinger Liquid. \emph{Phys. Rev. Lett.} \textbf{2013}, \emph{111}, 046402.
\bibitem{Dorner1} Dorner, R.; Clark, S.R.; Heaney, L.; Fazio, R.; Goold, J.; Vedral, V. Extracting Quantum Work Statistics and Fluctuation Theorems by Single-Qubit Interferometry. \emph{Phys. Rev. Lett.} \textbf{2013}, \emph{110}, 230601.
\bibitem{Mazzola1} Mazzola, L.; Chiara, G.D.; Paternostro, M. Measuring the Characteristic Function of the Work Distribution. \emph{Phys. Rev. Lett.} \textbf{2013}, \emph{110}, 230602.
\bibitem{Kuic1} Kui\'{c}, D.; Knapp, A.;
\v{S}aponja-Milutinovi\'{c}, D. Finite Time Path Field Theory Perturbative Methods for Local Quantum Spin Chain Quenches. \emph{Universe} \textbf{2024}, \emph{10}, 384.
\bibitem{Dadic1} Dadi\'{c}, I. Out of equilibrium thermal field theories: Finite time after switching on the interaction and Wigner transforms of projected functions. \emph{Phys. Rev. D} \textbf{2000}, \emph{63}, 025011.
\bibitem{Dadic2} Dadi\'{c}, I.; Klabu\v{c}ar, D. Causality and Renormalization in Finite-Time-Path Out-of-Equilibrium $\phi^3$ QFT. \emph{Particles} \textbf{2019}, \emph{2}, 92.
\bibitem{Dadic3} Dadi\'{c}, I.; Klabu\v{c}ar, D.; Kui\'{c}, D. Direct Photons from Hot Quark Matter in Renormalized Finite-Time-Path QED. \emph{Particles} \textbf{2020}, \emph{3},~676.
\bibitem{Dadic4} Dadi\'{c}, I.; Klabu\v{c}ar, D. Neutrino Oscillations in Finite Time Path Out-of-Equilibrium Thermal Field Theory. \emph{Symmetry} \textbf{2023}, \emph{15},~1970.
\bibitem{Lieb1} Lieb, E.; Schultz, T.; Mattis, D. Two Soluble Models of an Antiferromagnetic Chain. \emph{Ann. Phys.} \textbf{1961}, \emph{16}, 407.
\bibitem{Fabio-book} Franchini, F. \emph{An Introduction to Integrable Techniques for One-Dimensional Quantum Systems}; Lecture Notes in Physics; Springer: Cham, Switzerland, 
2017; Volume 940.
\bibitem{Goussev1} Goussev, A.; Jalabert, R.A.; Pastawski, H.M.; Wisniacki, D.A. Loschmidt echo. \emph{Scholarpedia} \textbf{2012}, \emph{7}, 11687.
\bibitem{Damski1} Damski, B.; Rams, M.M. Exact results for fidelity susceptibility of the quantum Ising model: The interplay between parity, system size, and magnetic field. \emph{J. Phys. A Math. Theor.} \textbf{2014}, \emph{47}, 025303.
\bibitem{Lancaster} Lancaster, T.; Blundell, S.J. \emph{Quantum Field Theory for the Gifted Amateur}; Oxford University Press: Oxford, UK, 
2014. 
\bibitem{Coleman} Coleman, P. \emph{Introduction to Many-Body Physics}; Cambridge University Press: Cambridge, UK, 2015. 
\bibitem{Le Bellac} Le Bellac, M. \emph{Thermal Field Theory}; Cambridge University Press: Cambridge, UK, 1996.
\bibitem{Sacco1} Sacco Shaikh, D.; Sassetti, M.; Traverso Ziani, N. Parity-Dependent Quantum Phase Transition in the Quantum Ising Chain in a Transverse Field. \emph{Symmetry} \textbf{2022}, \emph{14}, 996.
\bibitem{Duan1} Duan, L.-M; Demler, E; Lukin, M.D. Controlling Spin Exchange Interactions of Ultracold Atoms in Optical Lattices. \emph{Phys. Rev. Lett.} \textbf{2003}, \emph{91}, 090402.
\bibitem{Porta2} Porta, S.; Gambetta, F.M.; Cavaliere, F.; Traverso Ziani, N.; Sassetti, M. Out-of-equilibrium density dynamics of a quenched fermionic system. \emph{Phys. Rev. B} \textbf{2016}, \emph{94}, 085122.
\bibitem{Porta3} Porta, S.; Gambetta, F.M.; Traverso Ziani, N.; Kennes, D.M.; Sassetti, M.; Cavaliere, F. Nonmonotonic response and light-cone freezing in fermionic systems under quantum quenches from gapless to gapped or partially gapped states. \emph{Phys. Rev. B} \textbf{2018}, \emph{97},~035433.
\end{thebibliography}
\end{document}